\documentclass[fleqn,usenatbib]{mnras}

\usepackage{newtxtext,newtxmath}

\usepackage[T1]{fontenc}

\DeclareRobustCommand{\VAN}[3]{#2}
\let\VANthebibliography\thebibliography
\def\thebibliography{\DeclareRobustCommand{\VAN}[3]{##3}\VANthebibliography}

\usepackage{graphicx}	
\usepackage{amsmath}	
\usepackage{mhchem}
\usepackage{subcaption}
\usepackage{hyperref}

\usepackage{xcolor}

\newcommand{\maria}[1]{#1}

\title[Hydrogenation of thioacetaldehyde in the ISM]{Thioacetaldehyde (\ce{CH3CHS}) on interstellar ices: a key molecule to unravel two chemical dichotomies in the ISM}

\author[M. Mallo et al.]{
M. Mallo$^{1}$\thanks{MM: maria.mallo@iff.csic.es},
M. Sanz-Novo$^{2}$,
M. Agúndez$^{1}$,
J. Cernicharo$^{1}$,
G. Esplugues$^{4}$,
V.M. Rivilla$^{3}$,
I. Jiménez-Serra$^{3}$,
\newauthor
C. Cabezas$^{1}$,
and G. Molpeceres$^{1}$\thanks{GM: german.molpeceres@iff.csic.es}
\\
$^{1}$Instituto de Física Fundamental, CSIC, C/ Serrano 123, 28006 Madrid, Spain\\
$^{2}$Center for Astrochemical Studies, Max-Planck-Institut f\"{u}r extraterrestrische Physik, Giessenbachstrasse 1, Garching bei Munchen, 85748, Germany. \\
$^{3}$Centro de Astrobiología, CSIC-INTA, Ctra. de Torrejón a Ajalvir km 4, 28850 Torrejón de Ardoz, Madrid, Spain \\
$^{4}$Observatorio Astronómico Nacional (OAN), Alfonso XII 3, 28014 Madrid, Spain
}

\date{Accepted \today. Received \today; in original form \today}

\pubyear{2026}

\begin{document}
\label{firstpage}
\pagerange{\pageref{firstpage}--\pageref{lastpage}}
\maketitle

\begin{abstract}
	Thioacetaldehyde (\ce{CH3CHS}), recently detected in TMC-1, has an abundance approximately 36 times lower than its oxygen analog, acetaldehyde (\ce{CH3CHO}). This makes the \ce{CH3CHS}/\ce{CH3CHO} pair the one with the largest column density difference among the detected oxygen/sulfur analogue pairs in this cloud. 
    We investigate the hydrogenation pathways of \ce{CH3CHS} to address two chemical dichotomies in the ISM: (i) the differenciation between \ce{CH3CHS} and \ce{CH3CHO},  and (ii) the apparent absence of both \ce{CH3CHS} in the G+0.693-0.027 molecular cloud and ethyl mercaptan (\ce{CH3CH2SH}), in TMC-1.
    Our results reveal a complex scheme that involves multiple competing reactions, highlighting an efficient sequence of consecutive hydrogenations that can lead to \ce{CH3CH2SH}. This finding suggests that the high S/O ratio observed for thioacetaldehyde in TMC-1 ($\sim$36), and even more pronounced in G+0.693-0.027 ($\geq$112), may result from its conversion via hydrogenation on the ice surface, contrary to the case of \ce{CH3CHO}, which is more resistant to that chemical process.
    The straightforward hydrogenation of \ce{CH3CHS} on ices, which can also take place even in the gas-phase at 150 K, provides a reliable explanation for its non-detection in G+0.693-0.027, where grain-surface chemistry is expected to play an important role, favoring the conversion of \ce{CH3CHS} into \ce{CH3CH2SH}, which is indeed detected in G+0.693-0.027. In contrast, TMC-1 represents a more pristine gas-phase environment, where grain-surface chemistry has a lower impact. Under these conditions, \ce{CH3CHS} can persist, while \ce{CH3CH2SH} remains undetected. Overall, our results show the entirely different reactivity that contributes to the chemical complexity of two of the largest interstellar sulfur factories. 
      
\end{abstract}




\section{Introduction} \label{sec:intro}

Recent advances in the sensitivity of modern radio telescopes have led to a substantial increase in the number of detected sulfur-bearing molecules in cold dense clouds. These detections include a wide variety of species with increasing complexity such as \ce{HCS}, \ce{HSC} \citep{agundez2018detection},  \ce{HS2} \citep{esplugues2025first}, \ce{HCCS}, \ce{H2CCS}, \ce{H2CCCS}, \ce{C4S}, \ce{C5S} \citep{cernicharo2021tmc}, \ce{HCSCN}, \ce{HCSCCH} \citep{cernicharo2021sulphur},  NCCHCS \citep{cabezas2024laboratory}, \ce{HOCS+} \citep{Sanz-Novo2024}, \ce{CH2CHCHS} \citep{cabezas2025discovery}, dimethyl sulfide (DMS, \ce{CH3SCH3}) \citep{sanz2025abiotic} and even sulfur-bearing cyclic hydrocarbons that have recently been identified \citep{araki2026}. The large number of S-bearing molecules detected in  dense clouds has allowed to investigate the behaviour of sulfur in these environments, finding remarkable differences with respect to oxygen. This dichotomy between S and O will be addressed during the present work. Nevertheless, the chemistry of sulfur in the ISM is far from being completely understood, particularly in dense interstellar media, as it is clearly shown by the missing sulfur problem (see, for example \citealt{ruffle1999sulphur}, \citealt{Vidal2017} and \citealt{laas_modeling_2019} for major modeling studies on the issue), i.e. the fact that the major sink of interstellar sulfur has not been characterized yet. The molecules detected in the ISM only account for less than the 5\% of the cosmic abundance of sulfur \citep{asplund2009chemical}, and the main sulfur reservoir in these environments is still a major discussion point of modern astrochemistry \citep{herath2025missing, miranzo2025prodige}

Thioacetaldehyde (\ce{CH3CHS}) has recently joined the growing inventory of sulfur-bearing hydrocarbons detected in the cold dark cloud TMC-1 \citep{agundez2025detection}. It is the sulfur analog of acetaldehyde (\ce{CH3CHO}), one of the most abundant interstellar complex organic molecules (COMs) detected in cold dense clouds \citep{matthews1985detection}. In astrochemistry,  COMs are commonly defined as carbon-containing molecules composed of 6 or more atoms \citep{herbst2009complex} and they are expected to play a fundamental role in prebiotic chemistry. The chemistry of acetaldehyde has been extensively studied, including its formation mechanism both in the gas phase \citep{Balucani2015a,vazart2020gas} and on icy grains \citep{jin2020formation, lamberts2019formation}; however, its formation on grain surface entails some challenges \citep{enrique2021theoretical}. Following its detection by \cite{agundez2025detection}, several formation mechanisms for \ce{CH3CHS} in TMC-1 were proposed, but they have not yet been investigated in detail. Recently, \cite{rani2026interstellar} showed the efficiency of a gas-phase top-down scenario (i.e. starting from larger precursor species) for its formation although further astrochemical modelling is needed to determine the impact of the different formation and destruction pathways of \ce{CH3CHS} in the ISM.


The hydrogenation of \ce{CH3CHO} has been investigated through laboratory experiments and theoretical studies \citep{bisschop2007h, molpeceres25} due to its central role as a prototypical COM in the ISM. Recently it has been reported to proceed with small percentage of conversion to products other than \ce{CH3CHO} itself \citep{molpeceres25}, allowing to reconcile the chemistry of \ce{CH3CHO} with its isotopologues observations \citep{jorgensen_alma-pils_2018}. In the present work, we investigate the hydrogenation of \ce{CH3CHS} on icy grain surfaces in search of a particular behavior similar to that of \ce{CH3CHO}. A different chemistry between these species poses one of the two conundrums that we face in this work, since one of the main differences between sulfur and oxygen chemistry relies on the abundances of the S- and O-bearing molecules in the ISM. A clear dependence of the abundances with the degree of hydrogenation is observed, where the sulfur containing molecules with low degree of hydrogenation are more abundant than their oxygen analogues while the opposite occurs for highly hydrogenated species, as presented in \cite{agundez2025detection}. For example, the abundances of the unsaturated S-bearing chains \ce{C2S}, \ce{C3S} and \ce{C5S} are significantly larger than their oxygen analogues \citep{cernicharo2021tmc}, but the opposite behaviour is found for the more saturated molecules such as \ce{HCCCHS}. 

\maria{The pronounced differences in the abundances of O- and S-bearing species are likely due to their distinct reactivity, which is related to their intrinsic physico-chemical properties. The lower electronegativity of sulfur compared to oxygen results in less polar C=S bonds than the corresponding C=O bonds, which contribute to the enhanced reactivity of sulfur species. Consistently, the calculated HOMO-LUMO gap is smaller for \ce{CH3CHS} than for \ce{CH3CHO} (0.238 vs. 0.340, at the DFT level employed throughout this work), which supports the higher reactivity of the former.}

Besides the already intriguing deviations in the S/O ratios of different molecules, the second riddle posed by the detection of \ce{CH3CHS} in TMC-1 is its non-detection towards other chemically rich regions of the ISM, like the Galactic Center (GC) cloud G+0.693-0.027 (hereinafter G+0.693), an usual target of molecular line surveys \citep{requena2008galactic, jimenez2020toward, rivilla2022molecular, colzi2022deuterium, zeng2023amides, san2024first}, where, by contrast, \ce{CH3CHO} is observed abundantly \citep{sanz2022toward}. In fact, the hydrogenated analogue of \ce{CH3CHS}, ethyl mercaptan (\ce{CH3CH2SH}), has been detected \citep{rodriguez2021thiols} in G+0.693 along with its isomer DMS (\ce{CH3SCH3}; \citealt{sanz2025abiotic}), but remains undetected in TMC-1. 

Our goal with this article is to jointly answer the two questions left opened by \ce{CH3CHS} detection in TMC-1. First, the origin of the abnormal S/O ratios of \ce{CH3CHS} in this cloud and second, link them to the different chemistry operating in G+0.693.  To achieve our goals and link our results to the different chemistry found for \ce{CH3CHO}, we study the hydrogenation, as the most frequent reaction of dust grains, of \ce{CH3CHS}. As we will show during the article, analyzing the chemical behavior of \ce{CH3CHS} towards hydrogenation on icy grain surfaces provides key insights into why it is detected in some environments and not in others, and serves as an explanation to the abnormal S/O ratios found in TMC-1. 

Our paper is structured as follows. Section \ref{sec:methodology} describes the computational methods employed to investigate the hydrogenation of \ce{CH3CHS} on grain surfaces. The results of our calculations are presented in Section \ref{sec:results}, where we analyze the reaction pathways and the corresponding rate coefficients. In Section \ref{sec:discussion} we highlight the most significant findings and discuss their implications for the chemistry of sulfur-bearing molecules in the ISM, as well as the observational data related to the detection of \ce{CH3CHS} and \ce{CH3CH2SH} in TMC-1 and G+0.693, respectively. Finally, a brief summary is provided in Section \ref{sec:conclusions}.

\section{Methodology} \label{sec:methodology}

We investigate the hydrogenation of \ce{CH3CHS} using density functional theory (DFT) based calculations on water clusters serving as interstellar ice analogues. The choice of the DFT functional and basis set for the rest of the work is determined after benchmarking (details in Appendix \ref{appendix:benchmark}) different methods against DLPNO-CCSD(T) \citep{purvis_full_1982, dlpno, guo_communication_2018} (using a \texttt{Normalpno} localization scheme) calculations in the gas phase. Additionally, DLPNO-CCSD(T) was also tested against its non-local version (CCSD(T)) with the same basis set to ensure faithful representation of the electronic energies. After benchmarking, the method that was selected is M062X-D3/ma-def2-TZVP \citep{m06, d3, schafer1994fully, weigend2005balanced, weigend2006accurate} corrected with DLPNO-CCSD(T)/jun-cc-pV(T+d)Z \citep{papajak2011} single-point energies to refine the energetics of the stationary points. The combination of geometry optimizations at the DFT level with high-level single-point energy calculations is commonly used to balance accuracy and computational cost, and it is referred to in the text as DLPNO-CCSD(T)/jun-cc-pV(T+d)Z//M062X-D3/ma-def2-TZVP.	

To study the interaction of \ce{CH3CHS} with the ice surface, we created a cluster model made of 20 water molecules to simulate a local water ice arrangement. The construction of this ice model is done with our \textit{in-house} ice sampling tools relying on the atomic simulation environment (ASE) \citep{larsen2017atomic}. Briefly, we sequentialy shoot (with a kinetic energy pointing to the center of mass of the cluster of 30 K) randomly placed \ce{H2O} molecules, performing molecular dynamics (MD) simulations in the canonical ensemble for each trajectory and optimizing the structure every 5 water molecules being shot. A Langevin thermostat was employed to maintain the temperature constant at 30 K and the generic GFN-FF force field \citep{spicher2020robust} was used to perform a rapid initial exploration of the configurations. We built a total of 5 different water models with a different random seed, \maria{optimizing them at the M062X-D3/ma-def2-TZVP level of theory using tight convergence criteria to ensure a rigid ice model (attempting to avoid further relaxation once the adsorbate is placed)}. We use the \textsc{Orca} \citep{ORCA} electronic structure code throughout. Next, we created the ice-molecule clusters \maria{starting from the optimized ice structure} by placing the \ce{CH3CHS} molecule in different positions on a grid that spreads a spherical Fibonacci lattice deformed into an ellipsoid shape with a minimum distance of 4\r{A} between the adsorbate and the ice to avoid overlapping, in a similar manner as in \citet{molpeceres2021computational}. Ten configurations per ice were generated \maria{and the whole ice-molecule cluster was optimized}, leading to a total of 50 different initial geometries. The binding energy (BE) for each configuration is calculated as follows:
\begin{align}
     H_{\rm bin} = H_{\rm ice+ads} - (H_{\rm ice} + H_{\rm ads}), \label{eq:BE}
\end{align}
where $H_{\rm ice+ads}$ is the internal energy (enthalpies at 0 K) of the adsorbate-ice system, and $H_{\rm ice}$ and $H_{\rm ads}$ represent the internal energies of the isolated ice cluster and adsorbate, respectively. These internal energies are a sum of the electronic energy and the zero-point energy (ZPVE), all of them calculated at the M062X-D3/ma-def2-TZVP level. The binding energy distribution is used to select the most representative configurations for further study of the hydrogenation of \ce{CH3CHS} on the ice surface.

We investigated the hydrogenation of \ce{CH3CHS} on the ice surface for the binding site with the binding energy closest to the average value, after extracting a binding energy distribution (Section \ref{sec:first}). To characterize the stationary points of the reaction on the grain surface, we performed relaxed scan calculations with \textsc{Orca} at the M062X-D3 level, followed by transition state optimizations and intrinsic reaction coordinate (IRC) simulations to ensure that the transition states (TS) that define the energetic barriers are correctly described.
Once the reaction channels are characterized, we determine the hydrogenation rate coefficients using transition state theory (TST) corrected with tunneling effects by means of the Eckart model using the \textsc{Dl-Find} code \citep{kastner2009dl} interfaced with \textsc{Chemshell} \citep{metz2014c}, and employing an asymmetric form of the Eckart potential \citep{eckart1930penetration} \maria{to the entire ice-molecule cluster.} The use of the simplified Eckart model with respect to more sophisticated approaches like instanton theory or other methods explicitly including nuclear quantum effects in the derivation of the rate coefficient stems from the overall small activation energies for the hydrogen additions, in contrast with \ce{CH3CHO} for which the moderate activation energy made a more in detail treatment meritorious. \maria{The rate constants calculated for these reactions make use of the ``implicit surface approach'', that neglect the difference in the rotational partition functions between transition state and reactant \citep{Meisner2017}.}

A second hydrogenation on the ice surface of the main reaction products is also investigated to determine whether the efficient hydrogenation sequence of \ce{CH3CHS} can lead to the formation of more saturated species that could be locked into the grains, or by contrast return to the unsaturated \ce{CH3CHS} species is more likely. For this second stage, we use a similar protocol to the one shown in \citet{Molpeceres2021b, sanz-novo_conformational_2025}. Very briefly, a second set of hydrogen atoms are placed at a distance of 3 \text{\AA} from the center of mass of the preadsorbed molecule (the first hydrogenation product of \ce{CH3CHS}) forming a grid of 70 different configurations, with the additional constraint of a minimum distance of 2.5 \text{\AA} to any atom of the water cluster.  The system is allowed to relax from this initial set of positions, and the resulting structures are analyzed to determine the reaction channels and the approximated branching ratios. The biradical nature of the second reaction is ensured through a broken-symmetry DFT formalism but no coupled cluster refinement due to the lack of a proper energy reference. This lack of refinement is not expected to affect the results as in the case of the second hydrogenation we are only interested in a qualitative picture of the reaction channels and their relative importance.


\section{Results} \label{sec:results}

\subsection{First Hydrogenation} \label{sec:first}

\begin{figure}
    \centering
    \includegraphics[width=\columnwidth]{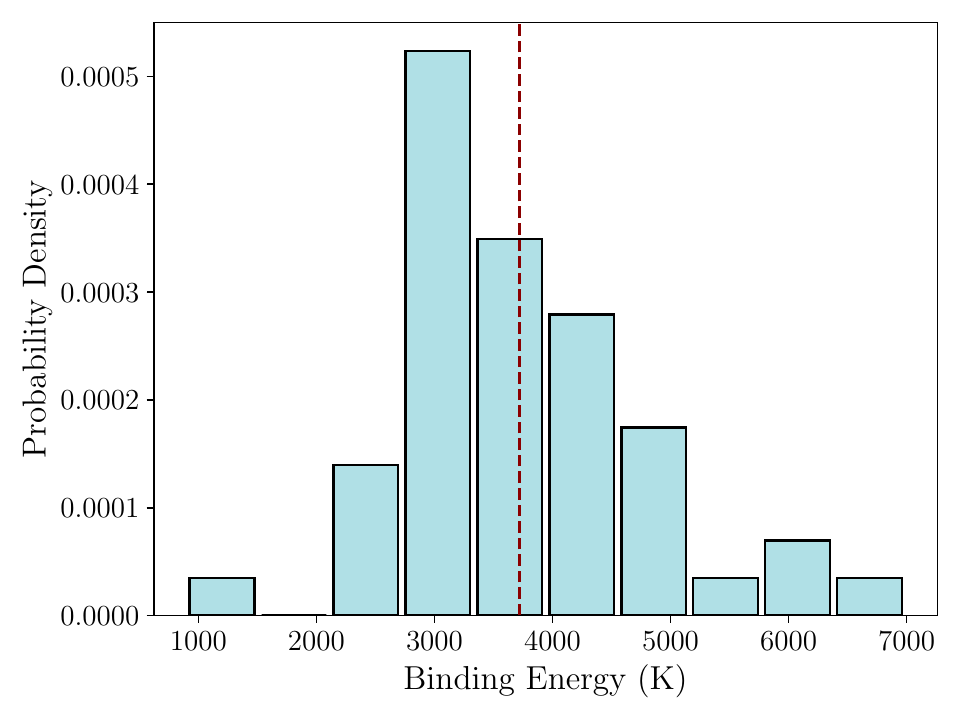} 
    \caption{Binding energy distribution of \ce{CH3CHS} on the ice cluster (including ZPVE correction). The average value is indicated with a dashed line.}
    \label{fig:BE}
\end{figure}

We begin our investigation by determining the water cluster binding site where the reactions will take place. The distribution of binding energies for the different thioacetaldehyde-water clusters is displayed in Figure \ref{fig:BE}. We observe a broad distribution of the energies, roughly following a Gaussian distribution centered around the model value of approximately 3000 K. The energies vary from 1000 K to 7000 K with an average value of 3724 K. 

This average binding energy is comparable to that of acetaldehyde, which stands at 3624 K (theoretical) and 3774 K (experimental) in \citet{molpeceres2022desorption}, and 4990 K in \citet{ferrero_acetaldehyde_2022}, 
although in the latter case there is a larger disparity of values owing to the strong O-H interactions on the surface. Nevertheless, the overall similarity in the binding energies between \ce{CH3CHS} and \ce{CH3CHO} suggests that the differential adsorption of the molecules onto the ice must not be a plausible explanation for the S/O partition in the gas-phase abundances. \maria{Our BE distribution does not preclude a more in depth study of the adsorption of \ce{CH3CHS} on the ice surface with a more sophisticated approach, such as in \citet{ferrero_acetaldehyde_2022}.}

\begin{figure}
    \centering
    \includegraphics[trim={0 5cm 0 3cm}, clip,width=0.8\columnwidth]{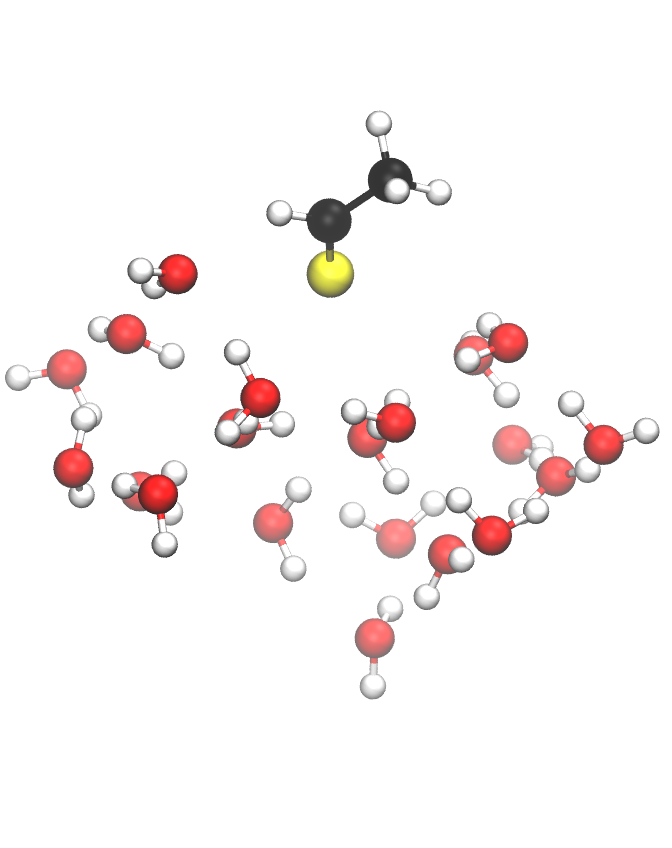} 
    \caption{Representation of the binding site of \ce{CH3CHS} on the ice cluster selected for the study of the hydrogenation. The color code is as follows: red for oxygen, white for hydrogen, yellow for sulfur and black for carbon.}
    \label{fig:cluster}
\end{figure}

The binding site chosen for studying the reactivity of \ce{CH3CHS} on the ice surface is the one with a binding energy of 3220 K, which is the closest to the average value of the distribution. The structure of the cluster with the adsorbed molecule is shown in Figure \ref{fig:cluster}.


\begin{figure*}
    \centering
    \includegraphics[width=0.9\textwidth]{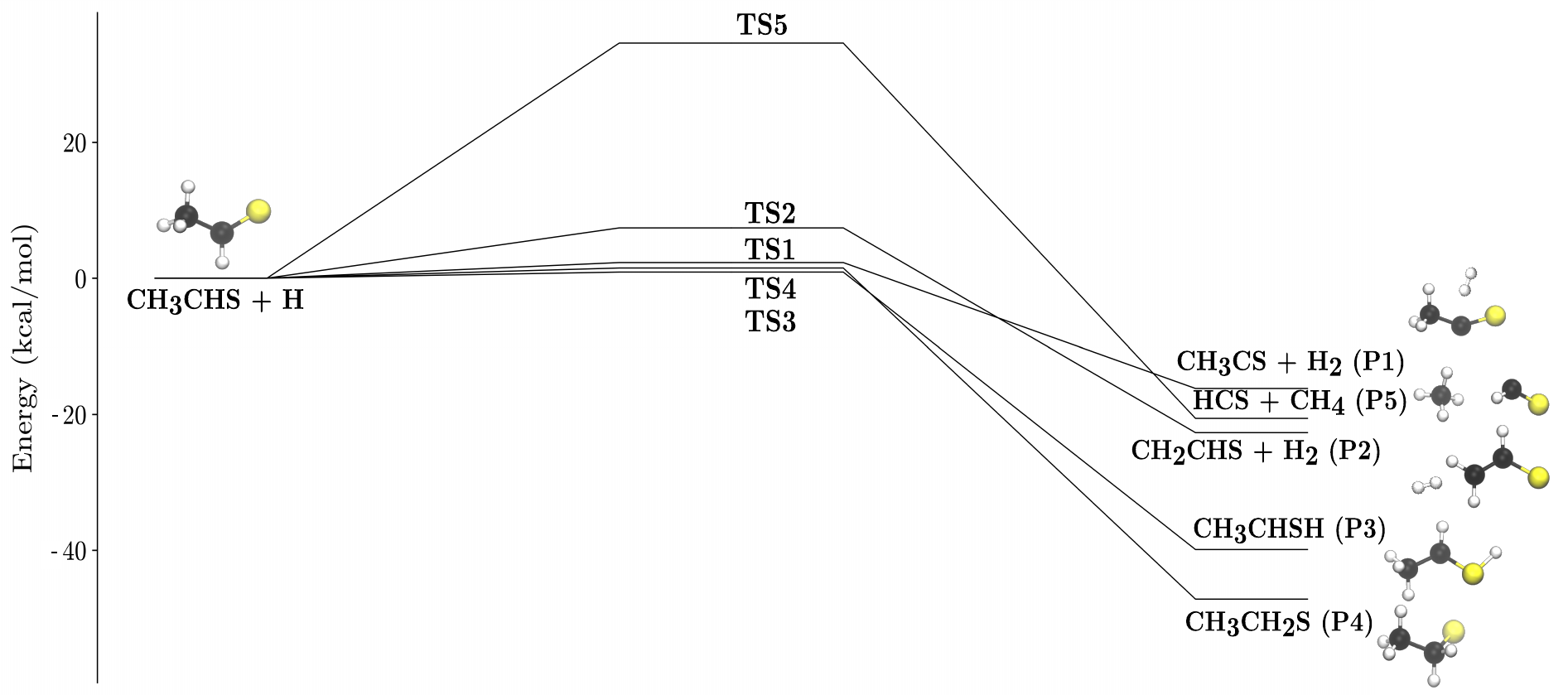} 
    \caption{Reaction pathways of the \ce{CH3CHS + H} reaction on the ice surface, including the schematic representation of reactant and products.}
    \label{fig:reac-path}
\end{figure*}

Following the reactivity scheme described in \cite{molpeceres25}, we evaluate all the possible reaction channels for the \ce{CH3CHS + H} reaction in order to properly study the hydrogenation and H-abstraction processes. 

\begin{align}
     \ce{CH3CHS + H} & \rightarrow \ce{CH3CS + H2} \label{eq:1} \tag{R1} \\
      & \rightarrow \ce{CH2CHS + H2} \label{eq:2} \tag{R2} \\
     & \rightarrow \ce{CH3CHSH} \label{eq:3} \tag{R3} \\
      & \rightarrow \ce{CH3CH2S} \label{eq:4} \tag{R4} \\   
      & \rightarrow \ce{HCS + CH4} \label{eq:5} \tag{R5} 
\end{align}

We begin our search of reactions (\ref{eq:1}-\ref{eq:5}) \maria{using the gas-phase model employed for the benchmark, meaning the isolated \ce{CH3CHS} molecule in the absence of explicit water molecules,} to assess the expected exothermicity and height of the barrier for the reactions. \footnote{\maria{It is worth noting that the gas-phase products are not necessarily the same as on the ice surface. The gas-phase model system is represented before any further evolution (e.g. collisional stabilization by third body, dissociation into bimolecular products, etc.)}}
We found that all the reactions are exothermic and proceed with a disparity of activation energies, from meager (Reactions \ref{eq:3} and \ref{eq:4}), moderate (Reactions \ref{eq:1} and \ref{eq:2}) or high (Reaction \ref{eq:5}), as shown in Table \ref{tab:energies_gas}.   
\begin{table}
    \centering
    \caption{Activation ($\Delta H^{\ddagger}$) and reaction energies ($\Delta H_{R}$) for the \ce{CH3CHS + H} reaction in the gas-phase model calculated at the DLPNO-CCSD(T)/jun-cc-pV(T+D)Z//M062X-D3/ma-def2-TZVP level. \maria{The analog results for \ce{CH3CHO} at the CCSD(T)/aug-cc-pVTZ//revDSD-PBEP86(D4)/jun-cc-pV(T+d)z level are included.} All of the energies are given in kcal mol$^{-1}$.}
\label{tab:energies_gas}
\begin{tabular}{ccccc}
\hline
& \multicolumn{2}{c}{\ce{CH3CHS + H}} & \multicolumn{2}{c}{\ce{CH3CHO + H}} \\
\hline
Reaction & $\Delta H^{\ddagger}$  & $\Delta H_{R}$  & $\Delta H^{\ddagger}$  & $\Delta H_{R}$   \\ 
    \hline
        R1  &  6.8  & -12.4 & 4.4 & -15.1 \\ 
        R2  &  7.5  &  -21.8 & 9.8 & -7.8 \\ 
        R3  &  1.6  &   -38.1 & 11.3 & -24.2 \\
        R4  &  3.5 & -45.9 & 7.4 & -14.4 \\
        R5  &  36.5 &   -19.5 & 34.1 & -20.7 \\
     \hline
    \end{tabular}
\end{table}


Once the reaction channels are characterized in the gas phase, we proceed to study them on the ice surface. The activation energies ($\Delta H^{\ddagger}$) and reaction energies ($\Delta H_{R}$) for the different reaction channels on the ice surface are summarized in Table \ref{tab:energies}. We find that the hydrogenation reactions on the ice surface follow the same general trends observed in the gas-phase exploration. However, the activation energies on the grain surface (Table \ref{tab:energies}) for most of the reaction pathways decrease with respect to those calculated for the gas phase (Table \ref{tab:energies_gas}). These interactions lead to slightly lower $\Delta H_{R}$ and $\Delta H^{\ddagger}$ on the ice surface than in the gas phase. Specially noteworthy in this context is the reduction of the $\Delta H^{\ddagger}$ for the H-abstraction reaction leading to \ce{CH3CS} (\ref{eq:1}). The reason for such a decrease is the stabilization of the transition state on the ice cavity through a pseudorotation. This requires a rearrangement of the water molecules and the \ce{CH3CHS} itself to accomodate the transition state, and subsequently the reactant state obtained through IRC calculations. This stabilization might not be as pronounced using a more rigid structure, e.g. different binding site or different ice model, and therefore there is a higher undertainty in the exact value of $\Delta H^{\ddagger}$ for \ref{eq:1} than for the rest of the reactions. Nevertheless, as we show in the next sections, our conclusions on the reactivity of \ce{CH3CHS} are robust against this magnitude.
 
We observe that the hydrogenation processes (\ref{eq:3} and \ref{eq:4}) have the smallest activation energies, with very low values of 0.9 and 1.5 kcal mol$^{-1}$, respectively, followed closely by \ref{eq:1} with a barrier of 2.3 kcal mol$^{-1}$ (see above). \ref{eq:2} shows a moderate barrier of 7.4 kcal mol$^{-1}$ and the fragmentation into heavy fragments (\ce{CH4 + HCS}, \ref{eq:5}) exhibits the highest barrier as predicted in the gas phase exploration, allowing us to discard this channel. In the case of \ce{CH3CHO} studied in \cite{molpeceres25}, the fragmentation into heavy elements (\ref{eq:5}) was also considered less relevant during the first hydrogenation. However, this preliminary exploration shows a notable contrast for Reactions \ref{eq:1}-\ref{eq:4} with respect to \ce{CH3CHO} \maria{(see Table \ref{tab:energies_gas})}. Overall, the different \ce{CH3CHO + H} pathways exhibit significantly higher barriers, suggesting that \ce{CH3CHO} is overall less reactive than \ce{CH3CHS}. Based on the energetic profile of both reactions, the H-abstraction reaction leading to \ce{CH3CO} (refered to as R1 in this work) is the most accesible pathway for acetaldehyde, while H-addition reactions, and in particular in the -S atom, sport the lowest $\Delta H^{\ddagger}$ the reactivity of \ce{CH3CHS}.

\begin{table}
    \centering
    \caption{Activation ($\Delta H^{\ddagger}$) and reaction energies ($\Delta H_{R}$) for the \ce{CH3CHS + H} reaction on the ice surface calculated at the DLPNO-CCSD(T)/jun-cc-pV(T+D)Z//M062X-D3/ma-def2-TZVP level. All energies are given in kcal mol$^{-1}$. \maria{The reaction energies correspond to the products still adsorbed into the grain surface, assuming no desorption into the gas-phase.}}
    \label{tab:energies}
    \begin{tabular}{ccc}
        \hline
    Reaction & $\Delta H^{\ddagger}$ & $\Delta H_{R}$  \\ 
        \hline
        R1  &  2.3  & -16.2 \\ 
        R2  &  7.4  & -22.7 \\ 
        R3  &  0.9  & -40.0  \\
        R4  &  1.5  & -47.2 \\
        R5  &  34.6  & -20.6 \\
        \hline
    \end{tabular}
\end{table}

The results from Table \ref{tab:energies} are illustrated in Figure \ref{fig:reac-path}. The most favorable pathways in the ice surface are the addition of the H atom to the C and S atoms of thioacetaldehyde to form the radicals \ce{CH3CHSH} (P3) and \ce{CH3CH2S} (P4), with barriers below 2 kcal mol$^{-1}$ pointing to an extremely easy hydrogen addition in comparison with the behavior found for \ce{CH3CHO} where H-abstraction was promoted. Nevertheless, energetic parameters alone are not sufficient to discard possible chemical pathways, so it remains to determine the actual value of the reaction rate coefficients, shown in Section \ref{sec:kinetic}.

\maria{To further explore the impact of the binding site on the reaction energetics and verify that our conclusions are not biased by the specific binding site chosen, we also investigated the most favorable forward pathway (\ref{eq:3}) using two other different binding sites, both of them being close to the average BE ($\sim$4000 K) but exhibiting apparent H-bonding of the S atom with the ice. In all cases, we find the formation of \ce{CH3CHSH} proceeding either with minimal barrier (0.9 kcal mol$^{-1}$) or even through submerged barriers (-0.05 and -0.54 kcal mol$^{-1}$, calculated at the DLPNO-CCSD(T) level). The presence of submerged barriers can be attributed either to the dual-level protocol applied in our calculations or real submerged behavior. Nevertheless, both binding sites show a difference of maximum 1.4 kcal mol$^{-1}$ with respect to the value reported in Table \ref{tab:energies} (0.9 kcal mol$^{-1}$), which is consistent with the uncertainty of the method, and most importantly, do not alter the diffusive rate constant that would be introduced into the models, as both cases will be totally dependent on the H diffusion rate than the specific reaction rate constants (diffusion controlled reaction).}

\subsection{Kinetic Analysis of the First Hydrogenation} \label{sec:kinetic}

\begin{table}
        \centering
        \caption{Unimolecular rate coefficients at 10 K for the different reaction channels of the \ce{CH3CHS + H} reaction calculated through TST corrected with tunneling effects using an Eckart model. The last column represents the absolute value of the imaginary vibrational frequencies ($\Omega$) of the transition states.}
        \label{tab:rate_constants}
        \begin{tabular}{ccc}
            \hline
        Reaction & k (s$^{-1}$) & $\Omega$ (cm$^{-1}$) \\  
            \hline
            R1  &  $1.6\times 10^{9}$  &  1810.7 \\ 
            R2  &  $3.0 \times 10^{4}$   & 1741.9 \\ 
            R3  &  $8.6 \times 10^{8}$  & 412.4 \\
            R4  &  $2.5 \times 10^{7}$ & 684.1\\
            \hline
        \end{tabular}
\end{table}

\begin{figure}
    \centering
    \includegraphics[width=\columnwidth]{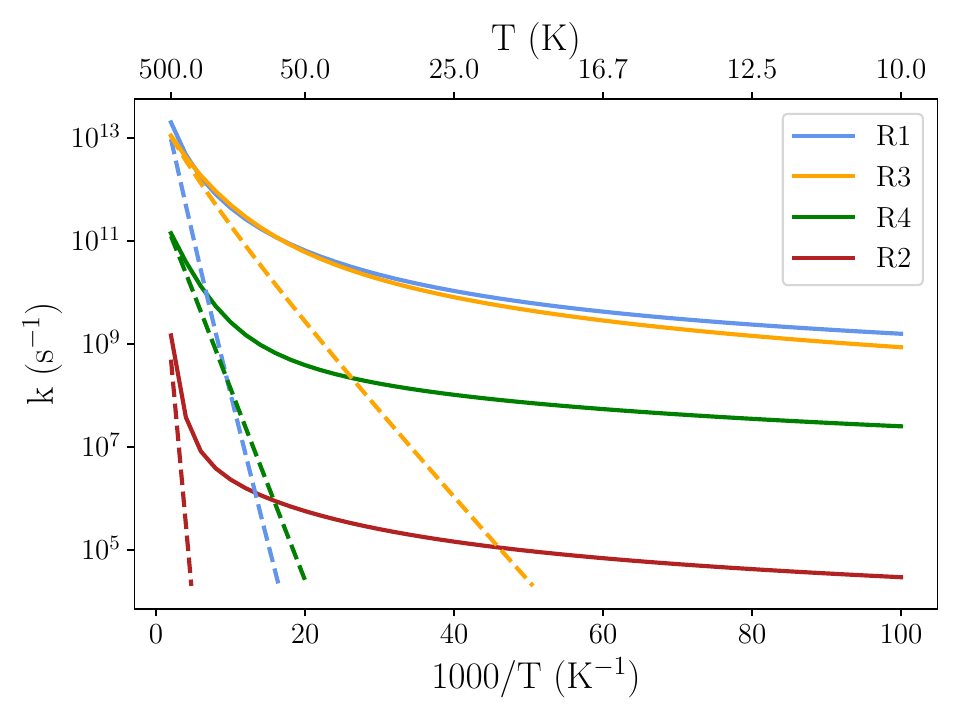} 
    \caption{Unimolecular rate coefficients in s$^{-1}$ for reactions R1-R4 as a function of temperature, from 500 K (left) to temperatures close to those found in the cold ISM (10 K, right). The real temperature (T) is represented in the upper axis for direct interpretation. \maria{Solid lines represent the rate coefficients including tunneling computed by means of Eckart model and dashed lines represent the rates without tunneling correction.}}
    \label{fig:rates}
\end{figure}

The unimolecular rate coefficients for the different reaction channels were calculated through TST corrected with tunneling effects by means of the (assymetric) Eckart model for the range of temperatures 10-500 K, although we note that \ce{CH3CHS} must desorb at much lower temperatures than 500 K. The evolution of the rate coefficients with temperature is shown in Figure \ref{fig:rates} and the values for the rate coefficients at 10 K are indicated in Table \ref{tab:rate_constants}. Since the outcome of the reaction is dominated by pathways \ref{eq:1}, \ref{eq:3} and (to a lower extent) by \ref{eq:4} and \ref{eq:2}, we can discard \ref{eq:5} from the analysis as it will not affect our conclusions. \maria{Figure \ref{fig:rates} also shows the rate coefficients without tunneling correction, which significantly drop with the decrease of temperature, reaching negligible values at low temperatures. This highlights the importance of tunneling effects in the hydrogenation of \ce{CH3CHS} on ice surface.}

Figure \ref{fig:rates} shows that H-abstraction to form \ce{CH3CS} (\ref{eq:1}) and  H-addition to form \ce{CH3CHSH} (\ref{eq:3}) dominate the kinetics over the whole temperature range, followed by \ref{eq:4}, with rates about two orders of magnitude lower. As shown in Table \ref{tab:rate_constants}, the reactions with lower $\Delta H^{\ddagger}$ also exhibit smaller $\Omega$. Both features typically imply higher tunneling probability.
At the temperatures closer to the ones in the cold regions of the ISM (10 K), the abstraction reaction \ref{eq:2} has a rate coefficient of about 10$^{4}$ s$^{-1}$, which makes it slower than hydrogen diffusion on interstellar ices \citep[spanning a wide range of diffusion coefficients centered around 10$^{5-6}$ s$^{-1}$][]{asgeirsson2017long,senevirathne2017hydrogen} and therefore, not a possible reaction channel. However, it could show significant variations in its rate coefficients if a more sophisticated approach for the description of quantum tunneling is used, owing to the higher tunneling efficiency (lower tunneling mass) of H-abstraction reactions. 
On the other hand, reactions \ref{eq:1}, \ref{eq:3} and \ref{eq:4} exhibit significantly higher rate coefficients than the typical values of diffusion, which means that all of them will be competitive. Nonetheless, for the purpose of our article, showing an effective pathway forward in the hydrogenation ladder (\ref{eq:3} and \ref{eq:4}), as opposed to the case found for \ce{CH3CHO} \citep{molpeceres25} is sufficient to prove the different chemical behavior of these species. This is the key distinctive behavior in the hydrogenation of \ce{CH3CHS} on ice grains, in contrast to that of \ce{CH3CHO}. While in the case of the latter molecule only H-abstraction is possible, for \ce{CH3CHS} both H-abstraction and H-addition are very efficient. Reaction \ref{eq:5} was discarded from our analysis.

In the gas phase, the forward hydrogenation of \ce{CH3CHS} is certainly governed by the addition of H to the sulfur atom, although  the final reaction products remain unknown in our investigation. Whereas grain surface kinetics have already been discussed above, analyzing the gas-phase pathways can provide additional insights into the destruction mechanisms of \ce{CH3CHS}. As shown in Table \ref{tab:energies_gas}, the presence of energy barriers exceeding 1 kcal mol$^{-1}$ suggest that the gas-phase hydrogenation is not efficient at the low temperatures of the ISM, although it could become more relevant at higher temperatures. \maria{To complete the kinetic analysis, we compute the bimolecular rate coefficients for the gas-phase reaction \ref{eq:3} using TST with quantum tunneling as described in Section \ref{sec:methodology}. It is worth noting that, in contrast with the calculations for the ice cluster model, we include rotational and translational degrees of freedom for the calculation of bimolecular rate constants. The \ce{CH3CHS} molecule and the H atom were treated as separated reactants, and the electronic energies of both of them were calculated at the DLPNO-CCSD(T) level and combined to define the reactant-state energy.} The results reveal a strong dependence of the rate with temperature, leading to promising implications for the detectability of \ce{CH3CHS} in regions with higher kinetic temperature, something that will be discussed in Section \ref{sec:discussion}. Specifically, the rate coefficient for reaction \ref{eq:3} is $1.2 \times 10^{-19}$ cm$^{3}$ s$^{-1}$ at 10 K, indicating a negligible contribution of this reaction under such conditions, although it increases significantly to $1.4 \times 10^{-13}$ cm$^{3}$ s$^{-1}$ at 150 K. As already mentioned, it is important to note that the gas-phase rate coefficient is provided without knowledge of the actual products of reaction, which are likely different from the ones found on the ice surface due to the lack of a third body to stabilize the products.

\subsection{Second Hydrogenation} \label{sec:second}

The theoretical investigation on the first hydrogenation of \ce{CH3CHS} on the ice surface shows a more complex chemistry than \ce{CH3CHO}, which was proven to be embeded in a H-abstraction and regeneration loop where no chemical evolution was observed \citep{molpeceres25}. In contrast, both addition and abstraction reactions can proceed efficiently for \ce{CH3CHS}.
These results indicate that an isoelectronic valence configuration is insufficient to extrapolate an equivalent radical chemistry within the same periodic group, showing that the reduced abundance of thioacetaldehyde may result from its efficient conversion into more saturated species through hydrogenation on ice grains. To further validate this hypothesis, we explore the subsequent hydrogenation steps of the addition products formed during the first hydrogenation stage (P1, P3 and P4), i.e. \ce{CH3CS + H}, \ce{CH3CHSH + H} and \ce{CH3CH2S + H}.

\begin{figure}
    \centering
    \begin{subfigure}{0.55\columnwidth}
        \centering
        \includegraphics[trim={1cm 3cm 0 1cm}, clip, width=\columnwidth]{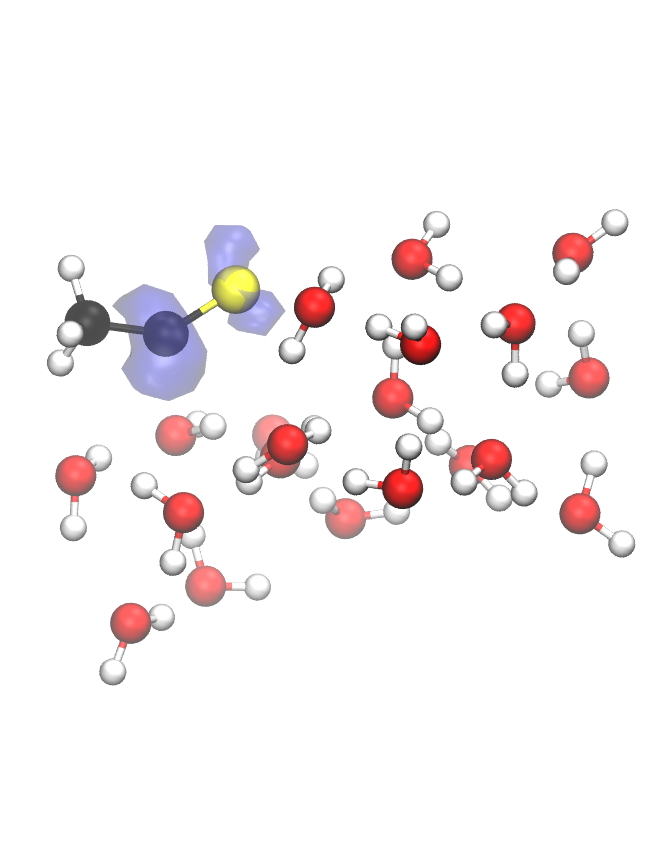} 
        \subcaption{\ce{CH3CS}, P1}
    \end{subfigure}
    \vspace{0.4cm}
    \begin{subfigure}{0.45\columnwidth}
        \centering
        \includegraphics[trim={1cm 1cm 0 1cm}, clip, width=\columnwidth]{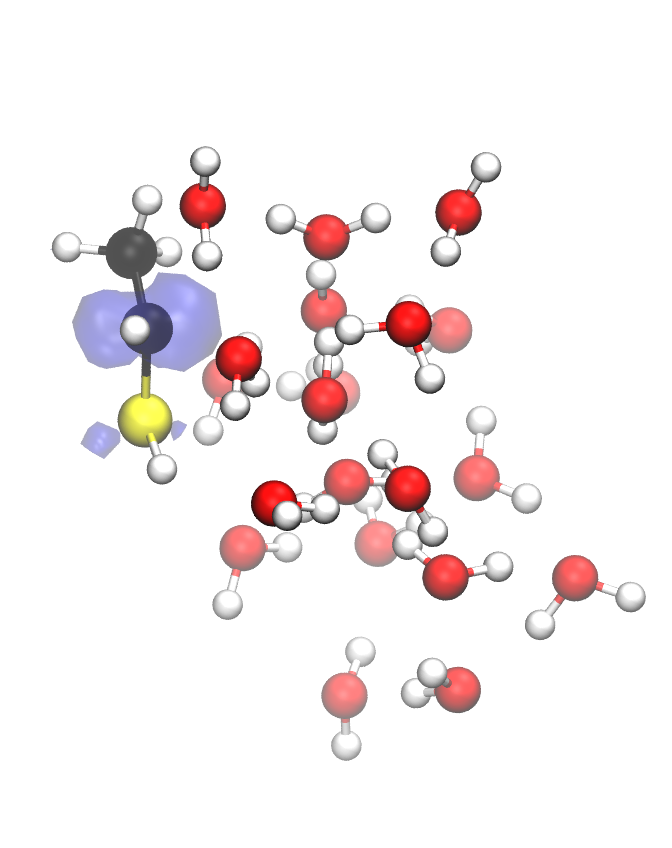} 
        \subcaption{\ce{CH3CHSH}, P3}
    \end{subfigure}
    \hfill
    \begin{subfigure}{0.45\columnwidth}
        \centering
        \includegraphics[trim={1cm 1cm 0 1cm}, clip, width=\columnwidth]{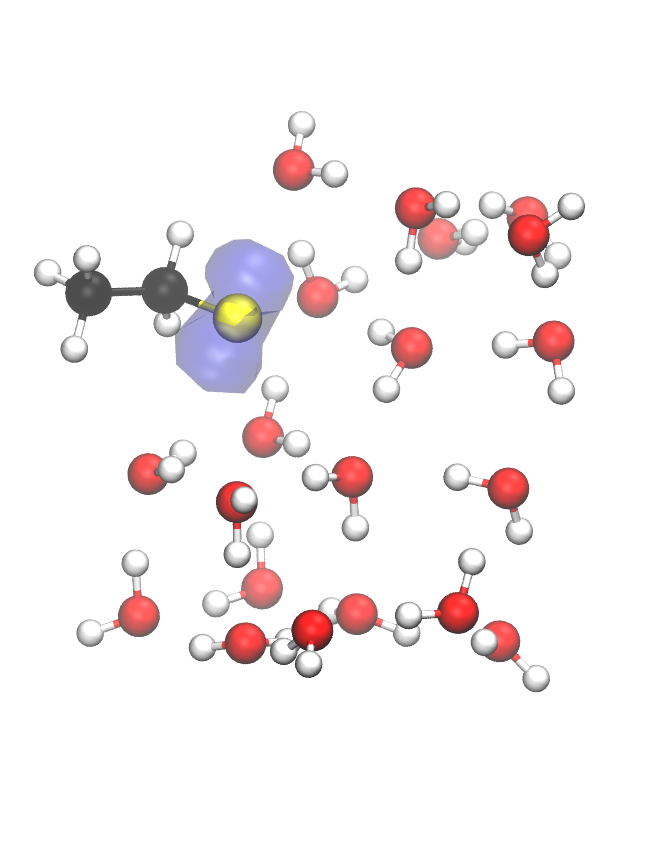} 
        \subcaption{\ce{CH3CH2S}, P4}
    \end{subfigure}
    \caption{Representation of the spin density distribution (in blue) of the products of the first hydrogenation. The isovalue is set to 0.01 a.u and only the spin density on the alpha channel is shown.}
    \label{fig:spin}
\end{figure}

Following the protocol for the investigation of radical-radical couplings presented in \citet{sanz-novo_conformational_2025} and briefly shown in Section \ref{sec:methodology} of this work, for each ice cluster (P1, P3 and P4) we generate 70 distinct configurations by randomly placing a hydrogen atom at a minimum distance of 3.0 \r{A} to any atom of the sulphur bearing radical.  Hydrogenations of these radicals species are expected to proceed without an activation barriers, at least in the positions with a higher spin-density (see Figure \ref{fig:spin}), yielding a more stable closed-shell product. Therefore, energy minimizations following a downhill path are performed for all the automatically generated configurations at the M062X-D3/ma-def2-TZVP level, e.g. without DLPNO-CCSD(T) refinement, using a broken symmetry formalism, to identify the most favorable chemical pathways. We note that, as expected, many initial configurations proved unreactive, which is something commonly observed when using this protocol \citep{sanz-novo_conformational_2025}, and can be explained as a combination of actual unreactive pathways or low tolerance of the geometry optimizer. The derived branching ratios are reported without considering the unreactive trajectories. The possible reaction pathways found in our search are summarized in equations \ref{eq:a}-\ref{eq:d}. 

\begin{align}
    \ce{CH3CS + H}  & \rightarrow \ce{CH3CSH} \label{eq:a} \tag{R\textsubscript{a}} \\
    \ce{CH3CHSH + H} & \rightarrow \ce{CH3CHSH2} \label{eq:b} \tag{R\textsubscript{b}} \\
     & \rightarrow \ce{CH3CH2SH} \label{eq:c} \tag{R\textsubscript{c}} \\
    \ce{CH3CH2S + H} & \rightarrow \ce{CH3CH2SH} \label{eq:d} \tag{R\textsubscript{d}} 
\end{align}

The outcomes of the second hydrogenation of P1, P3 and P4 (\ref{eq:a}-\ref{eq:d}) align with the expected reactivity based on the spin density distribution of the radicals displayed in Figure \ref{fig:spin}. In P1 and P3, the spin density is delocalized between the sulfur atom and the central carbon atom, although in P3 it is predominantly localized on the central carbon atom with a smaller contribution from sulfur. For both of them, the hydrogenation proceeds preferentially at the sulfur site, which can be attributed to the larger atomic size of sulfur and the S-H bond strength. In contrast, for P4 the spin density is solely localized on the sulfur atom, as confirmed by the exclusive hydrogen addition at this site.  
As outlined in Section \ref{sec:kinetic}, the following discussion will focus on the forward hydrogenation pathways (\ref{eq:b}-\ref{eq:d}), which are the most relevant for the chemical evolution of the system.

For the \ce{CH3CHSH + H} reaction, two possible products are formed: \ce{CH3CHSH2} and \ce{CH3CH2SH}, with branching ratios of 68.2\% and 31.8\%, respectively (\ref{eq:b} and \ref{eq:c}). Although \ce{CH3CHSH2} is not spectroscopically characterized, both $gauche$ and $anti$ isomers of \ce{CH3CH2SH} have been identified and targeted in searches across several sources \citep{kolesnikova2014spectroscopic,rodriguez2021thiols} (see Appendix \ref{appendix:observations}).
Additionally, \ce{CH3CH2SH} is the only product formed via \ce{CH3CH2S + H} (\ref{eq:d}). These hydrogenation reactions are barrierless and highly exothermic on the ice surface, and the reaction energies show differences in the thermodinamical stability of the products. \ce{CH3CH2SH} exhibits very strong coupling, with $\Delta H_{R}$ = -89.6  kcal mol$^{-1}$ when it is formed via reaction \ref{eq:b} and $\Delta H_{R}$ = -82.8 kcal mol$^{-1}$ via reaction \ref{eq:c}. \ce{CH3CHSH2} is less stable, with a reaction energy of -26.3 kcal mol$^{-1}$. 
Although the attack to the -SH moiety is the most favorable pathway in terms of branching ratio, the product is less stable and may undergo further transformations. We perform additional relaxed scan calculations to check the possible conversion of \ce{CH3CHSH2} into the more stable \ce{CH3CH2SH} via a proton transfer relay similar to the ones reported in \citet{Molpeceres2021c, molpeceres_carbon_2024}. 
Due to the complexity of this mechanism, which involves the concerted movement of several water molecules in the ice cluster, we use nudged elastic band (NEB) calculations \citep{henkelman2000climbing} using \textsc{Orca} to estimate the structure of the transition state at the M062X-D3/ma-def2-TZVP level. After locating the TS with the NEB method, IRC calculations were performed to confirm the connectivity of the stationary points. Our results indicate that the transformation proceeds through a small energy barrier of 2.2 kcal mol$^{-1}$ at the DLPNO-CCSD(T)/jun-cc-pV(T+D)Z//M062X-D3/ma-def2-TZVP level, which can readily be overcame on the ice surface.\footnote{Unlike the rest of reactions in this Section, the energies for the proton transfer reaction are corrected with local coupled cluster calculations, as it proceeds in a closed shell singlet PES.}  We compute the rate coefficient for this process using the methodology described in Section \ref{sec:kinetic}, yielding a value of $2.5 \times 10^{7}$ s$^{-1}$ at 10 K, supporting its viability. In the present study we do not calculate the binding energy of the final product on the ice surface (\ce{CH3CH2SH}) as it lacks of relevance for our particular investigation, but \cite{rani2026interstellar} report a value of 2430 K.

\begin{figure}
    \centering
    \begin{subfigure}{0.45\columnwidth}
        \centering
        \includegraphics[trim={1.5cm 0 1.5cm  0}, clip, width=\columnwidth]{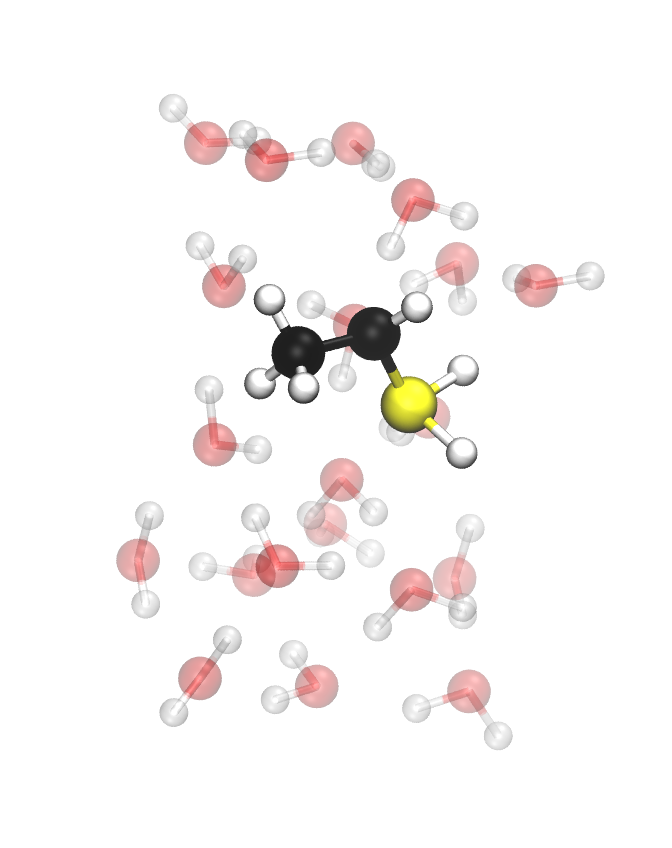} 
    \end{subfigure}
    \hspace{0.4cm}
    \begin{subfigure}{0.4\columnwidth}
        \centering
        \includegraphics[trim={1.5cm 0 1.5cm  0}, clip, width=\columnwidth]{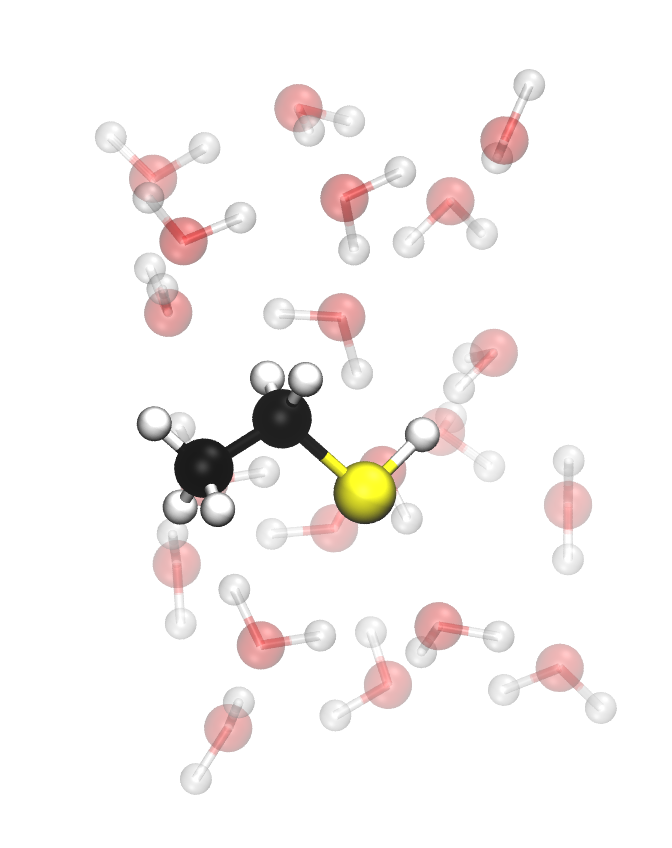} 
    \end{subfigure}
    \caption{Structures of the final hydrogenation products adsorbed on the ice surface: \ce{CH3CHSH2} (left) and \ce{CH3CH2SH} (right).}
    \label{fig:strucs}
\end{figure}

In addition to the energy minimization results, hydrogen abstraction from the main addition product (P3) leading to regeneration of the initial \ce{CH3CHS} was also investigated:
\begin{align}
    \ce{CH3CHSH + H} & \rightarrow \ce{CH3CHS + H2}, \label{eq:e} \tag{R\textsubscript{e}} 
\end{align}
through relaxed potential energy scan along the H-abstraction coordinate. \maria{This reaction proceeds through the formation of a pre-reactant complex (PRC) followed by a transition state at 0.84 kcal mol$^{-1}$ from the PRC at the M062X-D3/ma-def2-TZVP level. This process is highly exothermic, with a reaction energies of -60.9 kcal mol$^{-1}$. Consequently, the reverse reaction exhibits a large activation barrier, preventing an effective equilibrium between \ce{CH3CHSH + H} and \ce{CH3CHS + H2}.} These results highlight the broad reactivity of \ce{CH3CHS} on the ice surface. After undergoing either H-abstraction or H-addition, the species enters the reactive network and can experience subsequent hydrogenation or abstraction steps. Through this iterative sequence of reactions, the system can progressively evolve towards increasingly saturated intermediates until the final product, \ce{CH3CH2SH}, is ultimately formed.


\section{Discussion} \label{sec:discussion}

Beyond its chemical interest, the study of the hydrogenation of \ce{CH3CHS} on ice grains provides valuable insight into the sulfur-oxygen differentiation as a function of the molecular saturation, as previously discussed in \cite{agundez2025detection}. A direct analysis of the column densities of a series of sulfur and oxygen hydrocarbons in TMC-1 revealed that the abundance of sulfurated molecules decreases with the degree of hydrogenation. This pattern led to the conclusion that hydrogenation process were less efficient for S-bearing molecules, as suggested in that work. In the present case, a comprehensive analysis of the hydrogenation mechanism of \ce{CH3CHS} refutes the previous hypothesis, at least for the title molecule, and instead reveals a broad and diverse reactivity. The hydrogenation of \ce{CH3CHS} on the ice surface is shown to proceed efficiently, with high rate coefficients across the possible reaction channels. Furthermore, the evolution of this molecule toward more saturated species (\ce{CH3CH2SH}) is also demonstrated to be efficient and is supported by favorable energetic and kinetic parameters.

This behaviour contrasts with that of its oxygen analogue, which has been proven to suffer from limited conversion in \cite{molpeceres25}. For \ce{CH3CHO}, both experiments and calculations indicate that hydrogenation proceeds via H-abstraction followed by regeneration of the initial molecule, in a way that the molecule remains protected from permanent transformation. Besides, such a chemical cycle also significantly enhances chemical desorption \citep{Oba2018, Nguyen2021,molpeceres_formic_2025}, providing a pathway for the release of gas-phase \ce{CH3CHO}, where the molecule is detected through radio astronomical techniques. The disparity between the hydrogenation mechanisms of sulfur- and oxygen-bearing molecules offers an explanation for the observed S/O differentiation in the abundances of these species not only in TMC-1 but also in G+0.693 (see Appendix\,\ref{appendix:observations}). While the observed \ce{CH3CHS} abundance comes from its formation in the gas-phase \citep{agundez2025detection, rani2026interstellar}, the abundance of \ce{CH3CHO} is explained as a sum of the formation and limited depletion. It would remain to explain whether the same behavior is expected for other aldehyde (=O) thione (=S) pairs, like acroleine (\ce{CH2CHCHO}) and its sulfur analogue (propenethial; \ce{CH2CHCHS}), both detected in TMC-1 \citep{agundez_o-bearing_2021, cabezas2025discovery}. Besides, some extragalactic studies point the metallicity of the region as a determining factor for the S/O differenciation \citep{goswami2024contribution}, which could be addressed in future works.

An interesting exercise is to contextualize this reaction at the crossroad between two chemically rich, albeit significantly different molecular clouds, namely TMC-1 and G+0.693. 
\ce{CH3CH2SH} is not detected in TMC-1, with an upper limit to its column density of 9.4\,$\times$\,10$^{10}$ cm$^{-2}$ (see Appendix\,\ref{appendix:observations}). This, together with the fact that \ce{CH3CHS} is detected in that source with a column density of 9.8\,$\times$\,10$^{10}$ cm$^{-2}$ \citep{agundez2025detection} implies that the \ce{CH3CH2SH}/\ce{CH3CHS} abundance ratio in TMC-1 is $<$\,1. In sharp contrast, \ce{CH3CHS} is not detected in G+0.693, deriving an upper limit to its column density of $\leq$4.5\,$\times$\,10$^{12}$ cm$^{-2}$, which implies a \ce{CH3CHO}/\ce{CH3CHS} abundance ratio of $\geq$112 (Appendix\,\ref{appendix:observations}). However, \ce{CH3CH2SH} is indeed present in G+0.693 \citep{rodriguez2021thiols}, yielding a lower limit to the \ce{CH3CH2SH}/\ce{CH3CHS} abundance ratio of $\geq$9. These observational results point to a clear chemical differentiation between the two clouds.

The non-detection of \ce{CH3CHS} towards G+0.693 can then be understood in terms of the different chemical and physical conditions of both sources. TMC-1 is thought to be a molecular cloud with an estimated age of $\sim$10$^{5}$ years \citep{wakelam2006effect} and characterized by low temperatures of about 10 K, with dust and gas temperature being coupled. In contrast, G+0.693 represents a more energetic environment, likely experiencing a cloud-cloud collision \citep{zeng2020cloud}, with an increased cosmic-ray ionization ray \citep{Goto2013, Sanz-Novo2024} leading to a higher abundance of H atoms impinging on grains \citep{goldsmith_h_2005}, and a regime in which gas and grain temperatures are decoupled, with dust temperatures around 20 K \citep{rodriguez2004iso} and gas temperatures up to 150 K \citep{ginsburg2016dense, krieger2017survey}. The strong shock activity, summed to a more active grain chemistry, leads to efficient sputtering of dust grains and ejection of the grain molecules into the gas phase, thereby enhancing the abundances of molecules that otherwise would be observed in hot-cores by several orders of magnitude \citep{requena2006organic}. Meanwhile, the gas-phase chemistry is dominant in TMC-1, there is a lower impact of the grain hydrogenation and the products of grain-surface reactions are less efficiently released to the gas-phase, although we note that there are non-thermal mechanisms operating in the chemistry of several molecules \citep{wakelam_efficiency_2021, molpeceres_formic_2025}. In summary, a richer picture of the chemistry of grains can be obtained from G+0.693 while TMC-1 should show a more pristine gas-phase inventory.

In this context, the grain-surface hydrogenation of \ce{CH3CHS}  becomes particularly relevant in G+0.693 due to the physical conditions discussed above, which favor the efficient conversion of \ce{CH3CHS} into \ce{CH3CH2SH}. This is consistent with the positive detection of \ce{CH3CH2SH} in this source \citep{rodriguez2021thiols}, whereas TMC-1 reflects the lower impact of grain-surface chemistry, allowing \ce{CH3CHS} to persist.
In addition, the kinetic analysis reveal that the gas-phase hydrogenation is not feasible at the low temperatures characteristic of TMC-1 (10 K) but becomes more efficient at the higher temperatures of G+0.693 (150 K), yielding a bimolecular rate coefficient on the order of 10$^{-13}$ cm$^{3}$ s$^{-1}$ at this temperature (Section \ref{sec:first}). While this value still indicates a relatively slow reaction, it suggests that the gas-phase reaction is kinetically viable and may have a moderate impact on the destruction of \ce{CH3CHS} in G+0.693.

To conclude, our findings align with those reported in \cite{rani2026interstellar}, where the authors offer a complementary hypothesis for the exclusive detections of \ce{CH3CHS} and \ce{CH3CH2SH} across different sources, as well as  the \ce{CH3CHS}/\ce{CH3CHO} differenciation. They investigate the potential formation of \ce{CH3CHS} via top-down mechanisms initiated by H-abstraction of \ce{CH3CH2SH}, both in gas-phase and ice-surface models. Similarly to our conclusions, the physical conditions of the different environments are proposed as the main factor behind the detectability of these molecules, which makes both studies complementary and mutually reinforcing.

\section{Conclusions} \label{sec:conclusions}

In this work, we have investigated the hydrogenation pathways of \ce{CH3CHS}, which is the sulfurated molecule that exhibits the lowest column density ratio relative to its oxygenated analogue in TMC-1, i.e., the largest discrepancy between the abundances of sulfur- and oxygen-bearing species for the series of structures \ce{H_{x}C1S}, \ce{H_{x}C2S} and \ce{H_{x}C3S}, according to the current inventory of detected molecules in TMC-1.
The main findings can be summarized as follows:

\begin{itemize}
    \item We provide the high-level binding energy of \ce{CH3CHS} (3724 K), which is very similar to the one of \ce{CH3CHO}, thus discarding the differential adsorption of these molecules on the ice surface as a plausible explanation for the S/O partition in the gas-phase abundances.
    
    \item We characterize the reaction pathways of the \ce{CH3CHS + H} reaction on the ice surface, showing that multiple channels, including H-abstraction and H-addition, are viable. We calculate the activation energies and rate coefficients at the low temperatures of the cold ISM to confirm the presence of a complex network of competing reactions. Regarding the forward hydrogenation pathway, the formation of the \ce{CH3CHSH} radical is identified as the most favored route. This intermediate can subsequently undergo additional hydrogenation towards ethyl mercaptan (\ce{CH3CH2SH}), which has been detected in G+0.693 but not yet observed in TMC-1. These results provide a plausible explanation for the mutually exclusive detections of \ce{CH3CHS} and \ce{CH3CH2SH} in these sources. In warmer, more chemically developed environments such as G+0.693, grain-surface chemistry becomes increasingly relevant, and the desorption mechanisms facilitate the release of the final hydrogenation product (\ce{CH3CH2SH}) into the gas phase, enabling its detection. However, under the cold conditions of TMC-1, grain-surface chemistry has a more limited impact on the observed molecular abundances and the less saturated species (\ce{CH3CHS}) are more likely to be detected. Additionaly, thioacetaldehyde is expected to survive in the gas-phase due to the inefficiency of the gas-phase hydrogenation at low temperatures.
    
    \item We provide stringent 3$\sigma$ upper limits to the column densities of \ce{CH3CH2SH} in TMC-1 and \ce{CH3CHS} in G+0.693, based on independent, highly sensitive Q-band surveys carried out with the Yebes 40m telescope (and complementary millimeter data for G+0.693). These results pinpoint a clear chemical differentiation between the cold dark cloud TMC-1 and the shocked GC cloud G+0.693. 
    
    \item The great disparity on the abundances of \ce{CH3CHS} and \ce{CH3CHO} in TMC-1 and in G+0.693 can be attributed to differences in both their destruction and formation mechanisms. For the \ce{CH3CHO + H} reaction on the ice surface, \ce{CH3CHO} remains protected from permanent transformation through an H-abstraction and addition loop \citep{molpeceres25}, a type of chemical cycle known to promote chemical desorption. In contrast, \ce{CH3CHS} may experience several possible transformations depending on the attack site, ultimately being efficiently converted into more saturated species via H-addition on the grains, remaining spectroscopically invisible within the ices.
   
\end{itemize}


\section*{Acknowledgements}

We acknowledge from the grants PID2024-156686NB-I00 and RYC2022-035442-I funded by MICIU/AEI/10.13039/501100011033 and ESF+. G.M and M.M also acknowledge the support from project 20245AT016 (Proyectos Intramurales CSIC).  This work is supported by ERC grant No. 101218790 (\textsc{Isocosmos}) funded by the European Union. Views and opinions expressed are however those of the author(s) only and do not necessarily reflect those of the European Union or the European Research Council Executive Agency. Neither the European Union nor the granting authority can be held responsible for them.
We acknowledge funding support from Spanish Ministerio de Ciencia, Innovaci\'on, y Universidades through grant PID2023-147545NB-I00. Based on observations carried out with the Yebes 40m telescope (projects 19A003, 20A014, 20D023, 21A011, 21D005, and 23A024). The 40m radio telescope at Yebes Observatory is operated by the Spanish Geographic Institute (IGN; Ministerio de Transportes y Movilidad Sostenible).
M.S.-N. acknowledges funding from the Marie Sklodowska Curie fellowship SOUL (project number 101272259) funded by the European Union. Views and opinions expressed are however those of the author(s) only and do not necessarily reflect those of the European Union or the European Commission. Neither the European Union nor the granting authority can be held responsible for them.
V.M.R. acknowledges support from  the grant RYC2020-029387-I funded by MICIU/AEI/10.13039/501100011033 and by "ESF, Investing in your future", and from the Consejo Superior de Investigaciones Cient{\'i}ficas (CSIC) and the Centro de Astrobiolog{\'i}a (CAB) through the project 20225AT015 (Proyectos intramurales especiales del CSIC); and from the grant CNS2023-144464 funded by MICIU/AEI/10.13039/501100011033 and by “European Union NextGenerationEU/PRTR”. I.J.-S., V.M.R., and M.S.-N., acknowledge funding from grant No. PID2022-136814NB-I00 from MICIU/AEI/10.13039/501100011033 and by “ERDF, UE A way of making Europe”. I.J.-S. also acknowledges funding from the ERC grant No. 101125858 (\textsc{OPENS}) funded by the European Union. We also acknowledge support from the CSIC ILINK project SENTINEL (ILINK23017).


\section*{Data Availability}

A Zenodo repository including the optimized geometries of the stationary points is available at \url{https://doi.org/10.5281/zenodo.22811126}.  Any other data can be obtained upon request.



\bibliographystyle{mnras}
\bibliography{example} 




\appendix

\section{DFT Benchmark} \label{appendix:benchmark}

As mentioned in Section \ref{sec:methodology}, different DFT functionals were benchmarked against CCSD(T) and DLPNO-CCSD(T) calculations to select the most suitable method to describe the energy barriers of the present work. Table \ref{tab:benchmark} shows the activation energies of reactions \ref{eq:1}-\ref{eq:5}.

\begin{table}
    \centering
    \caption{Activation energies ($\Delta H^{\ddagger}$) for the H addition and abstraction reactions of \ce{CH3CHS + H} calculated with different DFT functionals and compared with CCSD(T) and DLPNO-CCSD(T) reference values, all of them given in kcal/mol.}
    \label{tab:benchmark}
    \resizebox{\columnwidth}{!}{%
    \begin{tabular}{cccccccc}
        \hline
    Method & B3LYP-D4 & M062X-D3 & WB97M-D4 & MN15 & MPWB1K & CCSD(T) & DLPNO-CCSD(T) \\ 
    Basis set & \multicolumn{5}{c}{ma-def2-tzpv} & aug-cc-pvtz & jun-cc-pV(T+t)Z \\
    \hline
    Reaction & \multicolumn{7}{c}{$\Delta H^{\ddagger}$ (kcal/mol)} \\
        \hline
R1  &  0.7  & 8.1 & 6.9 & 7.7 & 7.0 & 8.2 & 8.5 \\ 
R2  &  2.2 & 8.7 & 7.9 & 8.3 & 7.5 & 8.9 & 9.2 \\ 
R3  &  -0.5 & 0.5 & 0.8  & 0.3 & 0.3 & 0.5 & 0.9 \\
R4  &  0.1  & 2.5 & 3.0  & 2.0 & 2.3 & 2.3 & 2.8 \\
R5  & 30.3 & 35.2 & 36.2 & 36.7 & 36.8 & 36.8 & 36.6 \\
        \hline
    \end{tabular}
    }
\end{table}

The CCSD(T) and DLPNO-CCSD(T) energies are computed over optimized geometries with the reference method, which is revDSD-PBEP86-D4/jun-cc-pV(T+t)Z \citep{grimme2011effect, santra_minimally_2019, papajak2011}. All of the energies are obtained without including ZPVE corrections to focus on the electronic energy description of the different methods. Functionals were chosen to cover a wide range of families: Becke's hybrid GGA B3LYP, Truhlar's hybrid meta-GGA M062X \citep{m06} and MN15 \citep{mn15} and Chai and Head-Gordon's range-separated hybrid meta-GGA WB97M \citep{wb97} and MPWB1K, including dispersion corrections through the D3 or D4 methods \citep{d3, d4}. Concerning the basis sets, the ma-def2-TZVP basis set is used for the DFT calculations, while aug-cc-pvtz and jun-cc-pV(T+t)Z \citep{woon1994gaussian, papajak2011} are used for the CCSD(T) and DLPNO-CCSD(T) calculations, respectively.

DLPNO-CCSD(T) shows a good agreement with the CCSD(T) reference values, althought it slightly overestimates the energy barriers (below 0.4 kcal/mol) for all reactions. Therefore, we select the reduced DLPNO approach instead of the full CCSD(T) to correct the energies throughout the work. Finally, we selected M062X because it shows the better computational cost/accuracy ratio, with deviations below 0.4 kcal/mol for reactions \ref{eq:1}-\ref{eq:4} and about 1.2 kcal/mol for \ref{eq:5}, which is still in good agreement.


\section{Non-detection of \ce{CH3CH2SH} in TMC-1 and \ce{CH3CHS} in G+0.693} \label{appendix:observations}

We searched for \ce{CH3CH2SH} in TMC-1 using the latest data of QUIJOTE \citep{Cernicharo2021_quijote}, which is a line survey carried out with the Yebes\,40m telescope in the Q band (31.0-50.3 GHz) toward the position of the cyanopolyyne peak of the cold dense cloud TMC-1. The rotational spectrum of the $gauche$ conformer of \ce{CH3CH2SH} has been taken from the CDMS database, where data is based on laboratory studies \citep{Schmidt1975,Kolesnikova2014,Muller2016}. For a rotational temperature of 9 K \citep{Agundez2023}, the lines predicted as most intense in the Q band are $a$-type rotational transitions with $K_a$\,=\,0, 1, concretely the 4$_{1,4}$-3$_{1,3}$, 4$_{0,4}$-3$_{0,3}$, and 4$_{1,3}$-3$_{1,2}$, lying at 39.7, 40.5, and 41.4 GHz, respectively. These lines should appear as doublets due to the splitting caused by tunneling between the two equivalent configurations, but none of them are detected. The most stringent upper limit to the column density is imposed by the non-detection of the 4$_{0,4}$-3$_{0,3}$ doublet. Assuming a rotational temperature of 9 K, equal to the gas kinetic temperature of TMC-1, a line width of 0.60 km s$^{-1}$, and a emission size with a diameter of 80$''$ \citep{Agundez2023,Cernicharo2023}, we derive a 3\,$\sigma$ upper limit to the column density of gauche \ce{CH3CH2SH} of 9.4\,$\times$\,10$^{10}$ cm$^{-2}$.

We have also searched for \ce{CH3CHS} toward the GC molecular cloud G+0.693-0.027 using a sensitive, unbiased spectral survey conducted with the Yebes\,40m telescope in the Q band (31.075–50.424 GHz), as well as millimeter-wave observations (71.8–116.7 GHz, 124.8–175.5 GHz) carried out with the IRAM\,30m telescope (details of the observations provided elsewhere; \citealt{Rivilla2023,Sanz-Novo2023}.). For the rotational spectroscopic data of the molecule, we used entry 60801 from the Lille Spectroscopic Database (LSD; \citealt{LSD2025}), which is based on \cite{Margules2020}. After importing the spectroscopy into the \textsc{Madcuba}\footnote{Madrid Data Cube Analysis on ImageJ is a software developed at the Centro de Astrobiología (CAB) in Madrid: \url{https://cab.inta-csic.es/madcuba/}; see \citep{Martin2019}} package \citep{Martin2019}, we used the Spectral Line Identification and Modeling (SLIM) tool of \textsc{Madcuba} to generate the synthetic spectra of \ce{CH3CHS} assuming Local Thermodynamic Equilibrium (LTE) conditions. We used the same physical parameters derived for $gauche$ \ce{CH3CH2SH} in this source (i.e., a rotational temperature of 10 K, FWHM = 20 km s$^{-1}$ and $v$$_{lsr}$ = 69 km s$^{-1}$; \citealt{rodriguez2021thiols}) and compared the LTE model with the observed spectrum. We note that, in contrast to what was observed in TMC-1 \citep{agundez2025detection}, in this case the fine structure arising from the presence of a CH$_3$ internal rotor, which splits the rotational energy levels into the A and E substates, cannot be fully resolved due to the typical broad FWHM of the molecular line emission measured toward G+0.693 (FWHM $\sim$ 15$-$20 km s$^{-1}$; \citealt{requena2006organic,Zeng2018}). After inspecting the data, we did not identify any spectral features belonging to \ce{CH3CHS}. We thus derived a $3\sigma$ upper limit to its column density of $\leq 4.5 \times 10^{12}$ cm$^{-2}$ using the 4$_{1,3}$-3$_{1,2}$ A and E transitions (autoblended), located at 45.1111 and 45.1135 GHz, respectively, which are the lines predicted to be the brightest, falling within a clean region of the $Q$-band data. This translates into a relative abundance with respect to H$_2$ of 3.3 $\times$ 10$^{-11}$, adopting a $N$(H$_{2}$) = 1.35$\times$10$^{23}$ cm$^{-2}$ from \citet{martin_tracing_2008}. This value implies a \ce{CH3CHO}/\ce{CH3CHS} abundance ratio of $\geq$112 in G+0.693, which is significantly higher than general O/S trend observed in this source for related species (e.g., \ce{CH3OH}/\ce{CH3SH}, \ce{CH3CH2OH}/\ce{CH3CH2SH}, \ce{CH3OCH3}/\ce{CH3SCH3} abundance ratios of $\sim$31, $\sim$15 and $\sim$30, respectively; \citealt{Sanz-Novo2024,sanz2025abiotic}) and also deviates from the O/S solar value ($\sim$37; \citealt{asplund2009chemical}).

\bsp	
\label{lastpage}
\end{document}